\def\cm{cm$^{-1}$}
\def\pco{Pr$_2$CuO$_{4}$}
\def\DSco{DyScO$_{3}$}

\documentclass[prb,superscriptaddress,twocolumn,showpacs,amsmath,amssymb]{revtex4}
\usepackage[dvips]{graphicx}
\usepackage{graphicx}
\usepackage{color}

\begin{document}
\title{Optical study of superconducting Pr$_2$CuO$_{x}$ with $x\simeq 4$}

\author{G. Chanda}
\affiliation{Dresden High Magnetic Field Laboratory (HLD),
Helmholtz-Zentrum Dresden-Rossendorf, 01314 Dresden, Germany}
\affiliation{Institut f\"{u}r Festk\"{o}rperphysik, Technische
Universit\"{a}t Dresden, 01062 Dresden, Germany}
\author{R. P. S. M. Lobo}
\affiliation{LPEM, PSL Research University, ESPCI ParisTech, 10
rue Vauquelin, 75231 Paris Cedex 5, France} \affiliation{CNRS,
UMR8213, Paris, France} \affiliation{Sorbonne Universit\'es, UPMC
Univ. Paris 6, 75005, Paris, France}
\author{E. Schachinger}
\affiliation{Institute of Theoretical and Computational Physics,
NAWI Graz, Graz University of Technology, A-8010 Graz, Austria}
\author{J. Wosnitza}
\affiliation{Dresden High Magnetic Field Laboratory (HLD),
Helmholtz-Zentrum Dresden-Rossendorf, 01314 Dresden, Germany}
\affiliation{Institut f\"{u}r Festk\"{o}rperphysik, Technische
Universit\"{a}t Dresden, 01062 Dresden, Germany}
\author{M. Naito}
\affiliation{Department of Applied Physics, Tokyo University of
Agriculture and Technology, Naka-cho 2-24-16, Koganei, Tokyo
184-8588, Japan}
\author{A. V. Pronin}\email{artem.pronin@yahoo.com}
\affiliation{Dresden High Magnetic Field Laboratory (HLD),
Helmholtz-Zentrum Dresden-Rossendorf, 01314 Dresden, Germany}

\date{\today}

\begin{abstract}
Superconducting Pr$_2$CuO$_x$, $x\simeq 4$ (PCO) films with
$T^\prime$ structure and a $T_c$ of 27 K have been investigated by
various optical methods in a wide frequency (7 -- 55000 \cm) and
temperature (2 to 300 K) range. The optical spectra do not reveal
any indication of a normal-state gap formation. A Drude-like peak
centered at zero frequency dominates the optical conductivity
below 150 K. At higher temperatures, it shifts to finite
frequencies. The detailed analysis of the low-frequency
conductivity reveals that the Drude peak and a far-infrared (FIR)
peak centered at about 300 \cm persist at all temperatures. The
FIR-peak spectral weight is found to grow at the expense of the
Drude spectral weight with increasing temperature. The temperature
dependence of the penetration depth follows a behavior typical for
$d$-wave superconductors. The absolute value of the penetration
depth for zero temperature is 1.6 $\mu$m, indicating a rather low
density of the superconducting condensate.
\end{abstract}

\pacs{74.25.Gz, 74.25.nd, 74.72.Ek}

\maketitle

\section{Introduction}

\begin{figure}[t]
\centering
\includegraphics[width=7 cm,clip]{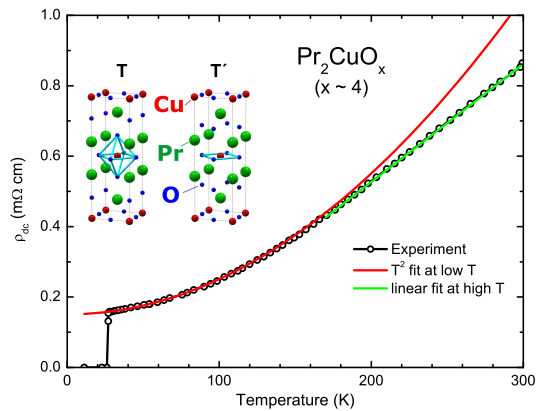}
\caption{(Color online) Temperature dependence of the in-plane dc
resistivity $\rho_{dc}$ of a MBE-grown $T^\prime$-PCO film [open
(black) circles] together with fits (lines) discussed in
Sec.~\ref{sec:res}. Schematic diagrams of the $T$ and $T^\prime$
structures are shown as an inset.} \label{rho}
\end{figure}

It is commonly accepted that the parent compounds of the
superconducting high-$T_c$ cuprates are antiferromagnetic
charge-transfer insulators and that superconductivity emerges upon
doping either with holes or electrons.\cite{Imada, ARM} There are
some similarities but also differences between hole- and
electron-doped cuprates. One similarity is that all cuprate
superconductors have a perovskite structure with the common
feature of square planar copper-oxygen planes separated by
rare-earth oxide (charge-reservoir) layers. On the other hand,
they differ in that the hole-doped cuprates have a $T$ structure
characterized by the presence of apical oxygen above and below the
CuO$_{2}$ planes, while the electron-doped cuprates have a
$T^\prime$ structure, where two sites are occupied by oxygen: O(1)
in the CuO$_{2}$ planes and O(2) within the rare-earth oxide
layers, with no apical oxygen located directly above the copper in
the CuO$_{2}$ plane, as shown in the inset of Fig. \ref{rho}. This
implies that the $T$ structure has six oxygen atoms, two of which
are in the apical positions, surrounding each copper (octahedrally
coordinated), while in the $T^\prime$ structure only four oxygens
surround each copper (square-planar coordinated).

There is also a large difference between the phase diagrams of hole-
and electron-doped cuprates. Whereas the antiferromagnetic phase
exists only over a small doping range (0 -- 4\,\%) in hole-doped
cuprates, it is more robust in electron-doped cuprates and persists
to higher doping levels (0 -- 11\,\%). Superconductivity occurs in
a doping range that is almost five times narrower for electron-doped
cuprates (11 -- 17\,\%) as compared to the hole-doped counterparts
(4 -- 32\,\%). While consensus on the phase diagram exists for the
hole-doped side, the situation for the electron-doped cuprates is
less obvious.

As early as in 1995, Brinkmann \textit{et al.}\cite{Brinkmann}
demonstrated that the superconductivity window in
Pr$_{2-x}$Ce$_x$CuO$_4$ single crystals can be extended down to a
doping level of 4\,\% by a special oxygen reduction and annealing
technique. Improved deposition and annealing techniques have
recently made it possible to produce thin films of electron-doped
parent compounds ($R_2$CuO$_4$, $R$ = Pr, Sm, Nd, Eu, and Gd) with
$T^\prime$ structure that, in fact, are metallic and
superconducting at low temperatures.\cite{Matsumoto1, Matsumoto2,
Matsumoto3, Matsumoto4, Yamamoto, Ikeda, YKrock1}

This sharp contradiction to earlier results is explained as being
due to the fact that although apical oxygen should not exist in the
ideal $T^\prime$ structure, in practice (especially in bulk samples)
it is usually not completely removed.\cite{YKrock1} This apical
oxygen in the $T^\prime$ structure acts as a very strong scatterer
and pair breaker.\cite{Sekitani} In contrast to bulk samples, the
large surface-to-volume ratio of thin films along with their tenuity
itself is advantageous in achieving the proper $T^\prime$ structure
with no apical oxygen.

The reported superconductivity in undoped cuprates puts a question
mark on the applicability of the charge-transfer-insulator picture
to electron-doped cuprates.\cite{Naito} Remarkably, recent
calculations on the basis of a newly developed first-principles
method show a radical difference between the parent compounds with
$T$ and $T^\prime$ structures.\cite{Das, Weber1, Weber3} The first
are found to be charge-transfer insulators, while the latter,
e.g., Pr$_{2}$CuO$_{4}$, are essentially metallic and their
apparent insulating nature may originate from magnetic long-range
order (Slater transition) which is competing with the metallic
ground state.\cite{Calder}

One should note, however, that it is still a question whether or
not $T^\prime$ superconductors are truly undoped or are still
doped by possible oxygen vacancies in the $R$O layers during the
reduction process. Since bulk $T^\prime$-$R_2$CuO$_{4}$
superconducting samples have not yet been synthesized, direct
measurements of the oxygen distribution are not available so far.
Nevertheless, neutron diffraction on Nd$_{2-x}$Ce$_{x}$CuO$_{4+y}$
single crystals shows that it is mostly apical oxygen which is
removed during reduction.\cite{Schultz, Radaelli} The synthesis of
bulk samples of a nominally undoped
$T^\prime$-(La,Sm)$_2$CuO$_{4}$,\cite{Ueda, Asai} and of heavily
underdoped
Pr$_{1.3-x}$La$_{0.7}$Ce$_{x}$CuO$_{4+\delta}$\cite{TAdachi} gives
hope that the oxygen stoichiometry might be determined in the near
future for this class of superconductors.

In this paper, we do not touch the issue of oxygen stoichiometry;
instead we present a comprehensive broadband optical investigation
of Pr$_2$CuO$_x$ (PCO) films with $x\simeq 4$. As argued above, it
is impossible to rule out doping by oxygen vacancies (if this is
the case, $x$ differs from 4 in our films). However, we will show
that our findings can also be consistently understood within the
picture, where superconductivity develops in undoped PCO (i.e.,
$x=4$). We demonstrate that the available PCO samples do show a
metallic as well as a superconducting optical response. We find
that this response can be reconciled with $d$-wave
superconductivity and the density of the superconducting
condensate is rather low. We do not observe any indication of a
normal-state pseudogap. All this supports ideas that the standard
charge-transfer-insulator picture might not be applicable to PCO.

\section{Experiment}

PCO films were grown by molecular beam epitaxy (MBE)
\cite{Yamamoto} on a (110)-oriented 0.35 mm thick \DSco{\rm}
substrate. The phase purity of these films was confirmed by x-ray
diffraction. The films were 100 nm thick with the $c$ axis
oriented perpendicular to the film's surface. Direct-current (dc)
resistivity was measured from 4 to 300 K by a standard four-probe
method.

Near-normal reflectivity from 40 to 55000 \cm{\rm} (5 -- 6800 meV)
was measured using a combination of two Fourier-transform
spectrometers (Bruker IFs113V and Bruker IFS66V/s) covering
frequencies from 40 to 22000 \cm{\rm} and a grating spectrometer
for room-temperature reflectivity measurements from 8000 to 55000
\cm. In order to obtain the absolute reflectivity of the sample,
we used an \textit{in situ} gold (for the infrared) or silver (for
the visible) overfilling technique.\cite{gold} With this
technique, we achieved an absolute accuracy in the reflectivity
better than 3 \% and the relative error between different
temperatures was of the order of 0.5 \%. The room temperature
reflectivity in the ultraviolet was measured against an aluminum
mirror and then corrected for the absolute reflectivity of
aluminum.

Normal-incident phase-sensitive transmission at 210 and 250 GHz (7
and 8.3 \cm{\rm}) was measured as a function of temperature with a
spectrometer employing backward-wave oscillators (BWOs) as sources
of coherent radiation.\cite{Kozlov} A Mach-Zehnder interferometer
arrangement of the spectrometer allows measurements of both the
intensity and the phase shift of the wave transmitted through the
sample. Using the Fresnel optical formulas for the complex
transmission coefficient of the two-layer system, the film's
complex conductivity as well as the penetration depth were
directly obtained from these measurements. This experimental
method has been previously applied to a large number of different
superconductors.\cite{Dressel} Technical details of our
experimental procedure can be found in Ref.~\onlinecite{Fischer}.

Optical properties of bare substrates were obtained from
measurements performed in the same frequency and temperature
windows as for the thin-film samples.

We investigated two thin films of PCO. The results, obtained on
the films, do not demonstrate any significant difference.
Hereafter we present results for one of the two films.

\section{Resistivity}
\label{sec:res}

Figure~\ref{rho} shows the temperature dependence of the
resistivity of the PCO film. The resistivity decreases
monotonically with decreasing temperature down to $T_c =$ 27 K.
The width of the superconducting transition is 0.8 K. The
temperature dependence of the resistivity can be described by the
power law
\begin{equation}
\rho(T) = \rho_{0} + AT^{n},
\end{equation}
with $\rho_{0}$ = 0.151 m$\Omega$cm, $A = 10^{-5}$
m$\Omega$cmK$^{-n}$, and $n = 2$ from $T_c$ up to 150 K. The
quadratic temperature dependence is in agreement with earlier
reports on superconducting Nd$_{2-x}$Ce$_x$CuO$_4$ films and
single crystals for temperatures below 200 K.\cite{Tsuei, Onose}
But, unlike Nd$_{2-x}$Ce$_x$CuO$_4$, where a slightly reduced
power law with $n$ ranging from 1.5 to 1.7 is observed above 200
K, we find a linear temperature dependence in PCO above 210 K. A
quadratic temperature dependence is often taken as evidence for
Fermi-liquid behavior.\cite{Abrikosov, Pines}

\section{Optical properties}

\subsection{Raw experimental data}
\label{subsec:a}

\begin{figure}[t]
\centering
\includegraphics[width=8 cm,clip]{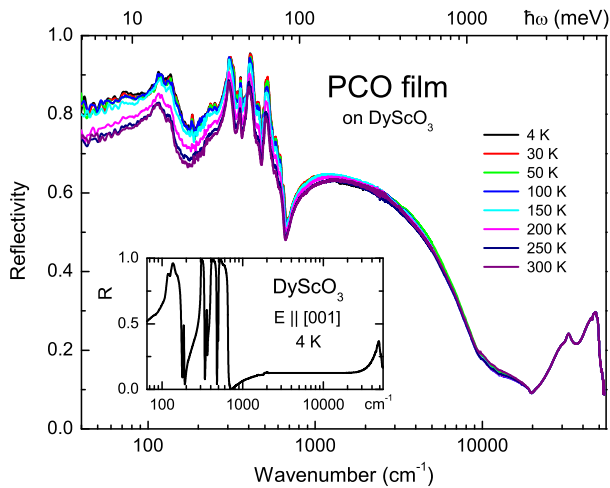}
\caption{(Color online) Reflectivity of the PCO thin film on a
\DSco{\rm} substrate as a function of frequency at various
temperatures listed in the legend. The \textbf{E} vector of the
probing radiation lies in the $ab$ plane of the film (and parallel
to the [001] axis of the substrate). The inset shows the
reflectivity of the bare substrate at 4 K.} \label{reflectivity}
\end{figure}

Figure~\ref{reflectivity} shows the as-measured in-plane
($ab$-plane) reflectivity of the PCO film on a \DSco{\rm}
substrate versus frequency at various temperatures. At low
frequencies, the reflectivity is quite high and increases with
decreasing temperature, typical for metals. A number of phonon
modes from the substrate and the film appears at frequencies below
700 cm$^{-1}$. The maxima seen above some 10000 cm$^{-1}$ can be
attributed to interband transitions.

\begin{figure}[b]
\centering
\includegraphics[width=4 cm,clip]{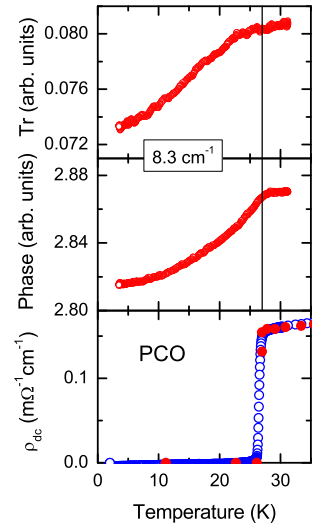}
\caption{(Color online) Examples of raw (i.e. not normalized to
the empty-channel measurements) phase-sensitive transmission
measurements at 8.3 cm$^{-1}$. Power transmission $Tr$ (top panel)
and phase shift (middle panel) of the wave passed through the PCO
film on the \DSco{\rm} substrate are shown as a function of
temperature together with a close-up of the dc resistivity
measurements around the superconducting transition (bottom panel).
The dc resistivity measurements were performed twice: on the fresh
film [solid (red) symbols] and after completion of all optical
measurements [open (blue) symbols]. The thin vertical line
indicates $T_{c}$.} \label{submm_raw}
\end{figure}

The changes to the reflectivity spectra induced by the
superconducting transition are not very well pronounced within our
experimental accuracy. This is because of a relatively high
transparency of the film. Thus, the results, obtained from the
reflectivity measurements, are only discussed in the normal state
in the course of the article.

The formation of the superconducting condensate can instead be
directly seen by use of our low-frequency phase-sensitive
transmission measurements. In Fig.~\ref{submm_raw} we present
examples of these measurements. The onset of the transition into
the superconducting state reveals itself immediately as a
reduction of the temperature-dependent power transmission $Tr$ and
the phase shift.\cite{Tr_phase} The penetration depth and the
superfluid density, obtained from these measurements, are
discussed in Sec. \ref{subsec:f}.

\subsection{Normal-state optical conductivity}
\label{subsec:b}

By applying a thin-film fitting procedure, described in detail in
App.~\ref{sec:ThinFlims}, we extract the film's complex optical
conductivity, $\sigma = \sigma_{1} + i\sigma_{2}$, from our
reflectivity spectra. Neither BWO data nor values of the dc
conductivity in the normal state have been utilized within this
fitting procedure.

\begin{figure}[t]
\includegraphics[width=8 cm, clip]{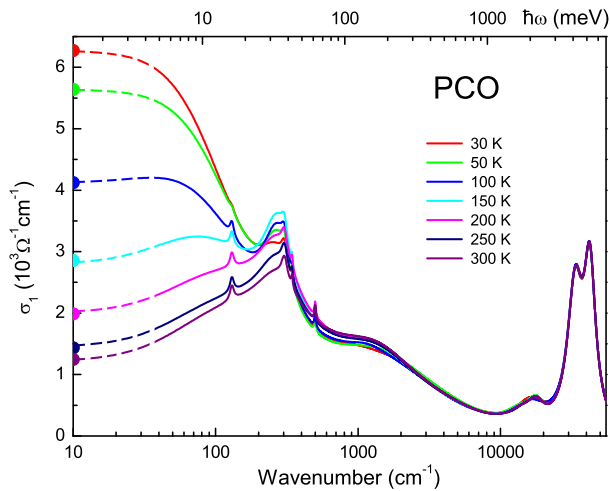}
\caption{(Color online) Real part of the optical conductivity of
PCO as a function of frequency for various temperatures listed in
the legend. Dots on the left-hand axis of the main panel represent
the dc-conductivity values.} \label{conductivity}
\end{figure}

The real part of the PCO optical conductivity obtained by this
modeling is shown in Fig.~\ref{conductivity} for various
temperatures indicated in the legend. As the lowest frequency of
the reflectivity measurements was 40 cm$^{-1}$ the data obtained
from this analysis below this threshold frequency are to be
considered as extrapolations and, thus, are shown as dashed lines.
Nevertheless, the zero-frequency limit of $\sigma_{1}$ evolves in
accordance with $\sigma_{dc}$ at all temperatures in the normal
state (bold points on the vertical left-hand axis of
Fig.~\ref{conductivity}).

\begin{figure}[b]
\centering
\includegraphics[width=\columnwidth, clip]{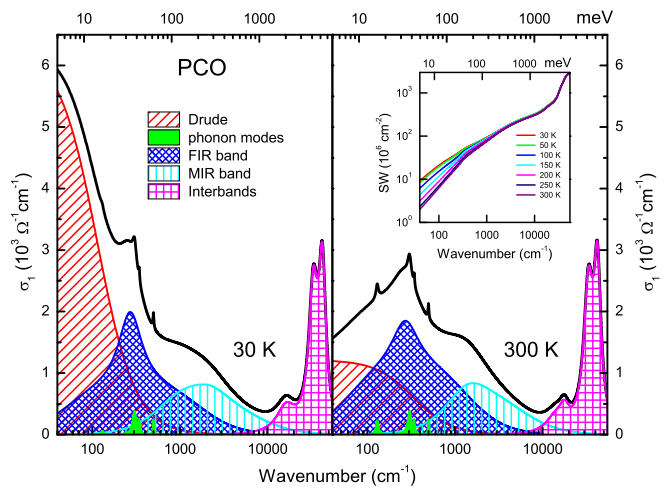}
\caption{(Color online) Decomposition of the real part of the
optical conductivity, $\sigma_1(\omega)$, at 30 K (left-hand
panel) and 300 K (right-hand panel). Inset: Frequency dependence
of the spectral weight of PCO as a function of the cutoff
frequency $\omega_c$ for various temperatures quoted in the
inset's legend.} \label{decomposition}
\end{figure}

At all temperatures above $T_c$, the optical conductivity of PCO
can be disentangled into a Drude component and a set of Lorentz
oscillators, representing a broad far-infrared (FIR) band, narrow
FIR peaks, a mid-infrared (MIR) band, and interband-transition
bands at the highest frequencies:
\begin{eqnarray}
\sigma(\omega) &=& \textrm{Drude + FIR band + FIR peaks +} \nonumber\\
&&\textrm{MIR band + interband transitions.} \label{decomp}
\end{eqnarray}
This becomes particularly evident from Fig.~\ref{decomposition}
where all these contributions are shown for 30 and 300 K. (We used
the Drude-Lorentz fitting procedure as described in
App.~\ref{sec:ThinFlims}. The FIR and MIR absorption bands have
been modeled with two Lorentzians each and we used three
Lorentzians for the interband transitions.)

We attribute the narrow and relatively weak peaks at 130 \cm{\rm},
304 \cm{\rm}, 343 \cm{\rm}, and 500 \cm{\rm} to infrared-active
phonon modes. Their frequency positions agree well with the
positions of strong phonon modes reported for nonsuperconducting
\pco{\rm} by Homes \textit{et al.}\cite{Homes}

It is worth noting here that in addition to the well pronounced
phonons, characteristic to the $T^\prime$ structure, other modes
which are not allowed by the crystal structure in the $T^\prime$
phase, have been observed by Homes \textit{et al.}\cite{Homes} The
authors have elaborated on the possible origin of these additional
modes but concluded eventually that some impurities and/or
contributions from different phases may play a role. This
conclusion is perfectly in line with claims made in Refs.
\onlinecite{Matsumoto1, Matsumoto2, Matsumoto3, Matsumoto4,
Yamamoto, Ikeda, YKrock1} that the complete removal of all apical
oxygen is extremely challenging and absolutely necessary for
superconductivity in undoped $T^\prime$ cuprates.

The bump at about 300\,\cm\ in the $\sigma_{1}(\omega)$ spectra
can be attributed to electron localization. Such behavior is known
for the superconducting cuprates\cite{Basov2} and is typical for
so-called bad metals where a certain degree of disorder is
inherently present.\cite{Emery, Mutou}

Using dynamical mean-field theory (DMFT) and iterated perturbation
theory, Mutou and Kontani demonstrated\cite{Mutou} that in a
strongly correlated metallic state, realized by a large on-site
repulsion energy, the optical conductivity develops a Drude peak
centered at $\omega=0$ at low temperatures and a shift of this
peak to finite frequencies above the Ioffe-Regel-limit temperature
$T_{IR}$, although the resistivity increases monotonically even at
$T > T_{IR}$. A temperature evolution of far-infrared
$\sigma_{1}(\omega)$, similar to the one observed here, has been
reported for underdoped Nd$_{2-x}$Ce$_x$CuO$_4$,\cite{Onose}
underdoped La$_{2-x}$Sr$_x$CuO$_4$,\cite{Takenaka} and zinc-doped
YBCO [YBa$_{2}$(Cu$_{1-x}$Zn$_{y})_{4}$O$_{8}$].\cite{Basov3}

The two highest-frequency absorption peaks at around 30000 --
40000 cm$^{-1}$ (4 -- 5 eV) are typical for the cuprates and
represent transitions into a band formed mostly by oxygen $p$
orbitals (see, e.g., Ref.~\onlinecite{Weber1}). The band slightly
below 20000 cm$^{-1}$ (e.g. around 1.5 -- 2 eV) is very similar to
the upper Hubbard band. Such absorption bands have been observed,
for example, in optical-conductivity studies of insulating undoped
Pr$_{2}$CuO$_{4}$\cite{Arima} and Nd$_{2}$CuO$_{4}$.\cite{Onose}
It is important to realize that the presence of such a band does
not necessarily require a charge-transfer gap. Moreover, LDA +
DMFT calculations\cite{Weber1} demonstrated that such a band may
perfectly coexist with a quasiparticle absorption peak (e.g. with
a metallic state) in the case of undoped Nd$_{2}$CuO$_{4}$ with a
perfect $T^\prime$ structure.

\subsection{Spectral weight}
\label{subsec:c}

We can trace the temperature evolution of each term in
Eq.~\eqref{decomp}. The mid-infrared band is almost invariant with
temperature, while the shape of the lower-frequency spectrum
(consisting of the Drude contribution and the FIR band) changes.
While at temperatures below 150 K the peak in $\sigma_{1}$ is
centered at $\omega = 0$ (the Drude term dominates), it shifts to
finite frequencies at $T > 150$~K. This shift is an indication of
the breakdown of the simple Drude-metal picture. It suggests a
continuous change in the charge transport from the low-temperature
coherent (Drude) to high-temperature incoherent
regimes.\cite{Lobo}

\begin{figure}[t]
\centering
\includegraphics[width=0.9\columnwidth, clip]{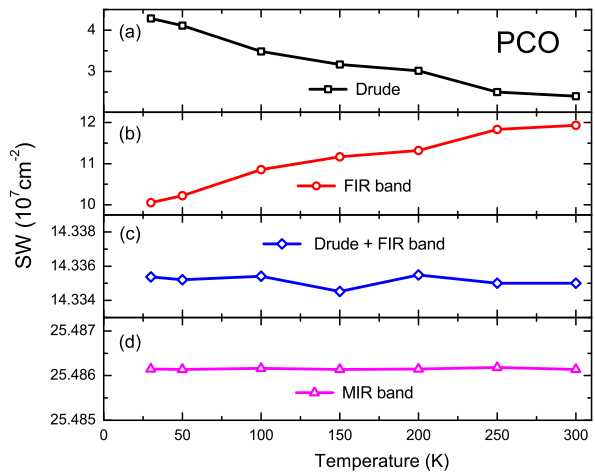}
\caption{(Color online) Temperature dependence of the spectral
weight: of the Drude term (a), the FIR band (b), the sum of the
two (c), and the MIR band (d) following the decomposition
according to Fig.~\ref{decomposition}} \label{sw_terms}
\end{figure}

In Fig.~\ref{sw_terms}, we plot the spectral weight (SW) of the
Drude term (panel a), the FIR band (panel b), the sum of the two
(panel c), and, for completeness, the MIR band (panel d). (The
spectral weight of each term in Eq.~\eqref{EpsDL} is just the
squared plasma frequency of the term.) As one can see from
Fig.~\ref{sw_terms}, the spectral weights plotted in panels (c)
and (d) are temperature independent, only the Drude and the
FIR-band spectral weights depend on temperature. It is obvious
that the spectral weight of the FIR band grows at the expense of
the Drude component with increasing temperature. We suggest that
this spectral weight transfer between the Drude and the FIR band
is related to the change in the transport properties, namely to
the change from the quadratic to the linear temperature dependence
of $\rho(T)$ which happens at comparable temperatures
(Fig.~\ref{rho}).

A qualitative picture of the spectral weight redistribution with
temperature in PCO can be studied by means of the total spectral
weight:
\begin{equation}
SW(\omega_{c}) =
8\int_{0}^{\omega_{c}}\!d\omega\,\sigma_{1}(\omega).
\end{equation}
It is plotted as a function of the cutoff frequency $\omega_c$ in
the inset on the right-hand panel of Fig.~\ref{decomposition}. At
low frequencies $\omega_c$, $SW(\omega_c)$ increases with
decreasing temperature, and it increases with $\omega_c$ for all
temperatures, finally developing an upturn around 10000 -- 15000
cm$^{-1}$. This upturn is due to interband transitions. Up to
15000 cm$^{-1}$, the spectral weight shows a temperature
dependence. Only at higher frequencies do the spectral-weight
curves merge, implying that above 15000 cm$^{-1}$ ($\sim1.9\,$eV)
the spectral weight is conserved as temperature changes. In other
correlated-electron materials, the spectral weight is known to be
conserved also only at frequency scales of a few eV.\cite{Imada,
Qazilbash} Thus, our results indicate the presence of electron
correlations in PCO.

In order to estimate the spectral weight and the plasma frequency
of the itinerant charge carriers only, we set $\omega_c / 2\pi=
9400$ cm$^{-1}$, thus cutting off the contribution from the
interband transitions. This gives a plasma frequency of 17700
cm$^{-1}$ (2.19 eV), a value comparable to those for other
high-$T_{c}$ cuprates. \cite{Onose, Uchida, Lee} Using the
relation between the charge-carrier density $n$ and the plasma
frequency ($\omega_{p}^{2} = 4\pi ne^2/m$), $n$ is estimated to
give $\sim 3.53 \times 10^{21}\,$cm$^{-3}$ assuming $m$ to be
equal to the free-electron mass $m_0$.

\subsection{Extended-Drude analysis}
\label{subsec:d}

To get further insight into the physics behind the optical
response of \pco{\rm}, we analyze the optical conductivity data in
terms of the extended (or generalized) Drude model which is widely
used for analysis of the optical properties of correlated electron
systems.\cite{JWAllen, Puchkov} The complex conductivity in this
model is given by
\begin{equation}
\sigma(\omega) =
\frac{1}{4\pi}\frac{\omega_{p}^{2}}{\Gamma(\omega)
-i\omega[1+\lambda(\omega)]},
\label{ext_Drude1}
\end{equation}
where $[1+\lambda(\omega)] = m^{*}(\omega)/m$ and
$\tau_{op}^{-1}(\omega) \equiv\Gamma(\omega)$ are the
frequency-dependent mass renormalization factor and the optical
scattering rate, respectively. Inverting Eq.~\eqref{ext_Drude1}
gives
\begin{equation}
1+\lambda(\omega) =
\frac{\omega_{p}^{2}}{4\pi}\frac{\sigma_{2}(\omega)}{\omega|\sigma(\omega)|^{2}};
\quad
\Gamma(\omega) =
\frac{\omega_{p}^{2}}{4\pi}\frac{\sigma_{1}(\omega)}{|\sigma(\omega)|^{2}}.
\label{ext_Drude2}
\end{equation}

The frequency-dependent optical scattering rate, obtained on the
basis of Eq.~\eqref{ext_Drude2} with $\omega_{p}$ = 2.19 eV, is
displayed in Fig. \ref{tau} as a function of frequency for various
temperatures listed in the legend. At $T < 150$~K, the general
trend in $\tau_{op}^{-1} (\omega)$ is to increase with frequency,
but this increase is nonmonotonic. This is due to phonons and the
localization mode discussed above. This mode reveals itself as a
bump at around 230 cm$^{-1}$ ($\sim$ 28 meV) in the optical
scattering rate. At $T > 150$~K, the scattering rate increases
rapidly as $\omega \rightarrow 0$. This is because at high
temperatures the localization mode dominates the Drude
contribution as it was discussed in relation to the
$\sigma(\omega)$ spectra.

\subsection{Eliashberg analysis and electron-boson spectral density}
\label{subsec:e}

The optical scattering rate is according to App.~\ref{sec:MaxEnt}
closely related to the electron-exchange boson interaction
spectral density $I^2\chi(\omega)$ [Eq.~\eqref{eq:InvTau}] which
is at the core of normal and superconducting state Eliashberg
theory.\cite{ESchach3} This theory can be applied to calculate
various normal and superconducting state properties and,
consequently, it is of quite some interest to gain knowledge on
$I^2\chi(\omega)$ by inverting $\tau^{-1}_{op}(\omega)$. This will
allow a more detailed analysis of our experimental results.

\begin{figure}[tb]
\centering
\includegraphics[width=8 cm,clip]{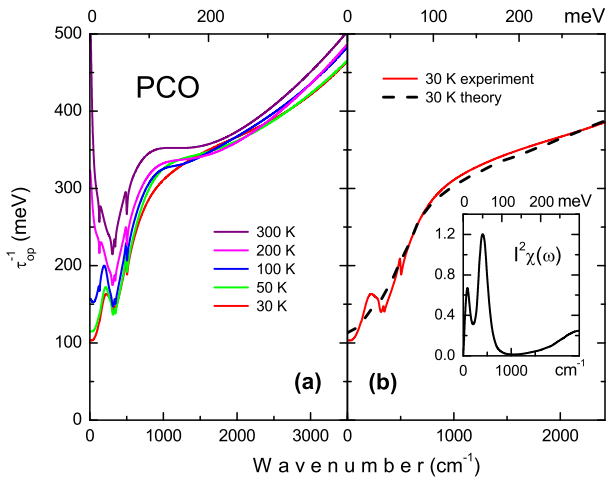}
\caption{(Color online) Panel a: The experimental optical
scattering rate in PCO for various temperatures listed in the
legend. Panel b: The experimental $\tau^{-1}_{op}(\omega)$ for
$T=30\,$K [solid (red) curve] and the Eliashberg-theory result
[dashed (black) curve] with an impurity parameter $t^+=15\,$meV,
see text. Inset in the panel: The electron-boson spectral density,
$I^2\chi(\omega)$, at $30\,$K as a result of a straightforward
inversion of the experimental $\tau^{-1}_{op}(\omega)$.}
\label{tau}
\end{figure}

It was also demonstrated by Schachinger {\it et
al.}\cite{ESchach1} that any nonzero contribution to
$I^2\chi(\omega)$ at some energy $\omega$ will result in an
increase of the optical scattering rate. Consequently the bump
observed in the optical scattering rate of PCO (Fig.~\ref{tau}) at
around $230\,$\cm\ ($\sim28\,$meV) cannot be caused by
electron-exchange boson interaction and is, therefore, not part of
the conducting-electron background. Nevertheless, we concentrate
on the normal-state $T=30\,$K data and perform a straightforward
inversion using the maximum-entropy procedure outlined in
App.~\ref{sec:MaxEnt} by inverting Eq.~\eqref{eq:InvTau} together
with the kernel Eq.~\eqref{eq:Shulga}. As the temperature and
frequency independent impurity scattering rate
$\tau^{-1}_{imp}=2\pi t^+$ is not known, this is an iterative
process which is performed by slowly increasing $t^+$ until a
smooth function $I^2\chi(\omega)$ with no pronounced spikes in the
immediate vicinity of $\omega=0$ has been found. This resulted in
$\tau^{-1}_{imp}\sim100\,$meV ($t^+=15\,$meV) which is quite
substantial but in good agreement with what has been reported for
the system PCCO.\cite{ESchach} Furthermore, we restricted the
frequency range of the inversion to $\omega\in[0,300]\,$meV
because between $100\le\omega\le 300\,$meV,
$\tau^{-1}_{op}(\omega)$ develops only a moderate increase with
energy.

It has to be pointed out, though, that Eq.~\eqref{eq:Shulga} is
only approximate. Therefore, we use the spectrum $I^2\chi(\omega)$
which resulted from the inversion process to calculate the
quasiparticle self-energy using the full normal-state
infinite-bandwidth Eliashberg equations. The complex infrared
conductivity $\sigma(\omega,T)$ is then calculated using a Kubo
formula\cite{lee} and the resulting optical scattering rate is
calculated from Eq.~\eqref{ext_Drude2}. A comparison of this
result with the data requires some adaptation of the original
$I^2\chi(\omega)$ spectrum in order to achieve the best possible
agreement with the data. This final spectrum is presented in the
inset of Fig.~\ref{tau}. It shows a double-peak structure which is
followed by a deep valley and a hump at higher frequencies. The
low-energy peak is at $\sim 11\,$meV and the high-energy peak can
be found at $\sim 50\,$meV. Similar double-peak spectra have been
reported for PCCO by Schachinger {\it et al.}\cite{ESchach} and
for La$_{1.83}$Sr$_{0.17}$CuO$_4$ (a hole-doped cuprate) by Hwang
{\it et al.}\cite{Hwang} both with a less pronounced low-energy
peak. It is most likely that the bump around $\sim 28\,$meV in the
PCO $\tau^{-1}_{op}(\omega)$ data is responsible for this
overpronouncement of the low-energy peak in the PCO
$I^2\chi(\omega)$ spectrum.

We found that the mass renormalization factor, which can be
calculated as the first inverse moment of $I^2\chi(\omega)$, is
$\lambda=4.16$.

A comparison of theoretical and experimental
$\tau^{-1}_{op}(\omega)$ data for $T=30\,$K is presented in
Fig.~\ref{tau}(b). The solid (red) curve represents the data while
the dashed (black) curve presents the result of our theoretical
calculations on the basis of the $I^2\chi(\omega)$ spectrum shown
in the inset. (Of course, good agreement between theory and data
cannot be expected in the energy region around the bump at $\sim
28\,$meV.)

\begin{figure}[t]
\centering
\includegraphics[width=\columnwidth,clip]{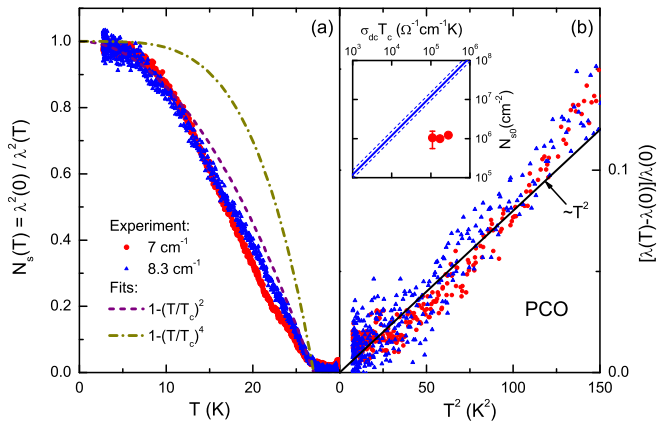}
\caption{(Color online). Panel (a): Superfluid density, $N_s(T) =
\lambda_L^2(0)/\lambda_L^2(T)$, as a function of temperature.
Panel (b): Low-temperature variation of the normalized London
penetration depth, $[\lambda_L(T)-\lambda_L(0)]/\lambda_L(0)$, as
a function of temperature squared. Data derived from the
millimeter-wave conductivity measurements at $7\,$\cm and
$8.3\,$\cm are presented by solid (red) circles and solid (blue)
triangles, respectively. Panel (a) contains for comparison the
temperature dependence $N_s(T) = 1-(T/T_c)^2$ [thin dashed
(purple) line] which is expected for a nodal superconductor and
$N_s=1-(T/T_c)^4$ [thin dashed-dotted (olive) line] for a fully
gapped superconductor. In panel (b) a quadratic power law of the
reduced penetration depth is indicated by a thin solid (black)
line. Inset: A universal relation between the zero-temperature
superfluid density, $N_{s0}$, and the product of normal-state dc
conductivity and $T_{c}$, as found in Ref.~\onlinecite{Homes1}
[straight solid (blue) line], reported ``error bars" of this
relation [straight dashed (blue) lines], and the data obtained for
three PCO films investigated in this study and in
Ref.~\onlinecite{Pronin} [bold (red) circles]. The error bars are
shown for the least accurate data.} \label{lambda}
\end{figure}

\subsection{Penetration depth and Superfluid density}
\label{subsec:f}

The temperature dependence of the penetration depth of PCO was
obtained experimentally by means of phase-sensitive
millimeter-wave measurements.\cite{Kozlov} Using the Fresnel
optical formulas for the complex transmission coefficient, the
in-plane complex conductivity of the film was calculated directly
from the measured transmission coefficient and phase shift. The
penetration depth was then calculated from $\sigma_2$ by using
$\lambda_{L}=c/(4\pi\omega\sigma_{2})^{1/2}$, where $c$ is the
vacuum speed of light and $\omega$ is the frequency of the
incoming radiation. We found $\lambda_{L}(T\rightarrow0) \approx
1.6\pm 0.1$ $\mu$m.

Figure~\ref{lambda}(a) presents the normalized superfluid density
$N_s(T) = n_s(T)/n_s(0) = \lambda_L^2(0)/\lambda_L^2(T)$ measured
at $7\,$\cm\ [solid (red) circles] and $8.3\,$\cm\ [solid (blue)
triangles] \textit{vs} temperature. We added curves for $N_s(T) =
1-(T/T_c)^2$ [thin dashed (purple) curve] and $N_s(T) =
1-(T/T_c)^4$ [thin dashed-dotted (olive) curve] for comparison.
They are supposed to mimic the temperature dependence of $N_s(T)$
for nodal ($d$-wave) and fully gapped ($s$-wave)
superconductivity, respectively. Obviously, the former curve
describes the data reasonably well, whereas the fully gapped
behavior can certainly be ruled out.

More sensitive is the low-temperature variation of the penetration
depth $[\lambda_L(T)-\lambda_L(0)]/\lambda_L(0)$ as a function of
the square of the temperature. The data are presented in
Fig.~\ref{lambda}(b) using the same symbols as in
Fig.~\ref{lambda}(a). For a nodal superconductor a quadratic power
law [thin solid (black) line] is to be expected and the data are
in agreement with this power law at lowest $T$.

In returning to the absolute value of the zero-temperature
penetration depth [$\lambda_{L}(T\rightarrow0) \approx 1.6$] we
conclude that the density of the superfluid condensate is very low
in PCO as compared to typical values for optimally doped cuprates,
where the penetration depth is smaller by a factor of 5 to
10.\cite{Basov1} For example, in optimally doped PCCO $\lambda_{L}
= 330$~nm.\cite{Zimmers1} This points toward a doping-related
nature of superconductivity in our PCO samples, as large values of
$\lambda_L(0)$ are typical for either underdoped or overdoped
regimes. However, the value of $\lambda_L(0)$ found here is so
big, that within this picture our PCO sample must be far off
optimal doping. This is rather unlikely, because the critical
temperature of our film is definitely too high for a heavily
underdoped or overdoped sample. It is also important to note that
a possible degradation of $T_{c}$ while the optical measurements
have been performed can be excluded because the dc resistivity
measured after the completion of all optical measurements does not
differ from the resistivity measured on the fresh film (see bottom
panel of Fig.~\ref{submm_raw}).

Furthermore, we would like to note that our second PCO sample had
a $\lambda_L(0) = 1.5 \pm 0.1$ $\mu$m and a very similar value of
$T_{c}$. Another PCO film prepared by a different method (metal
organic decomposition \textit{vs} MBE for the current films) was
reported by some of us to have $\lambda_L(0) = 1.55 \pm
0.25$~$\mu$m and $T_c = 27.5$~K.\cite{Pronin}

It is revealing to note that all these samples do not at all fit a
supposedly universal relation reported by Homes \textit{et
al.}\cite{Homes1} which connects the superfluid density [or
$\lambda_L(0)$] of cuprates to the product of $T_{c}$ and
normal-state conductivity $\sigma_{dc}$. This relation was
obtained from an analysis of experimental data on doped samples,
e.g. on doped charge-transfer insulators, and works for all doping
levels. According to Zaanen\cite{Zaanen} the existence of this
universal relation reflects the fact that the normal state of the
doped cuprates is extremely viscous (dissipative).\cite{Zaanen1}
All the PCO samples studied here and in Ref.~\onlinecite{Pronin}
are far off this universal relation (see the inset in
Fig.~\ref{lambda}). It is tempting to explain this fact as a sign
of a possible departure from the charge-transfer-insulator picture
in PCO: while Homes' relation reflects the physics behind the
doped-insulator picture, it does not necessarily work any longer
whenever this picture loses its validity in cuprates.

Nevertheless, if our samples are indeed doped, it seems to be
reasonable to assume that they must be underdoped rather than
overdoped. This assumption is based on the method used to prepare
the samples (reduction of oxygen content), on the high value of
$T_{c}$, and on the low superfluid density.

\subsection{Absence of a pseudogap feature}
\label{subsec:g}

A well-known characteristic feature of the underdoped cuprates is
the occurrence of a pseudogap, i.e. a partial normal-state gap in
the electronic density of states. Such a pseudogap has been
observed in underdoped cuprates by many experimental
methods.\cite{Timusk} In optical experiments, the occurrence of a
pseudogap below a characteristic temperature can manifest itself
in different ways. In hole-doped cuprates, the pseudogap is seen
as a suppression of the low-frequency scattering rate.
\cite{Basov1} In electron-doped cuprates a suppression of the MIR
reflectivity is observed that corresponds to a reduced real-part
MIR optical conductivity and to a nonmonotonic behavior of the
so-called restricted spectral weight,
RSW$(\omega_{L},\omega_{H},T)$.\cite{Zimmers} Here, RSW is defined
as
\begin{equation}
RSW(\omega_{L},\omega_{H},T)=8\int_{\omega_{L}}^{\omega_{H}}
\!d\omega\,\sigma_{1} (\omega,T), \label{eq:4gc}
\end{equation}
where $[\omega_{L}$, $\omega_{H}]$ is the restricted frequency
range of interest.

\begin{figure}[t]
\centering
\includegraphics[width=7 cm,clip]{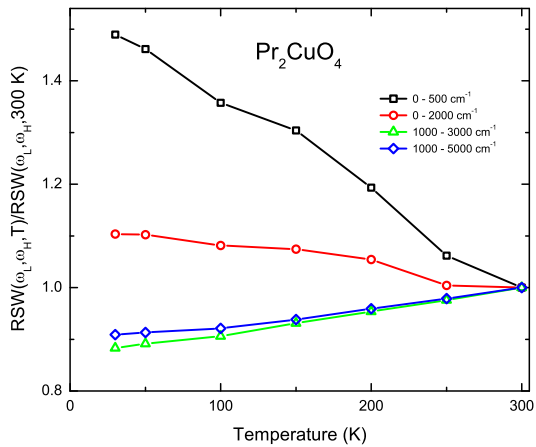}
\caption{(Color online). Temperature dependence of the restricted
normalized spectral weight (RSW) with the integration boundaries
[see, Eq.~\eqref{eq:4gc}] indicated in the legend.} \label{sw2}
\end{figure}

It follows from the reflectivity and conductivity data presented
in Figs.~\ref{reflectivity} and \ref{conductivity} that such a
normal-state gap is not evident in the current system. The absence
of the normal-state pseudogap is further confirmed by the results
presented in Fig.~\ref{sw2}. In this figure, the restricted
spectral weight $\textrm{RSW}(\omega_L,\omega_H,T)$ normalized to
the spectral weight at $T=300\,$K is plotted as a function of
temperature for four frequency ranges $[\omega_L,\omega_H]$ as
quoted in the legend. It is evident that the normalized restricted
spectral weight displays in all four cases a monotonic temperature
dependence. This rules out the existence of a normal-state
pseudogap.\cite{Zimmers} In addition, the optical scattering rate
(see Fig.~\ref{tau}) shows (apart from the low-frequency features
due to localization, phonon modes, and MIR bands) no
temperature-dependent suppression, that might (similarly to the
hole-doped cuprates) indicate a pseudo-gap-like
feature.\cite{Timusk}

Thus, we conclude that a pseudogap is absent in the PCO films, in
contrast to most underdoped high-$T_c$ cuprates. In our view, this
difference can be related to the absence of an antiferromagnetic
phase in PCO.\cite{YKrock1, Naito, YKrock, TAdachi} In
electron-doped cuprates, the magnetic order induces the pseudogap.
With doping, the Neel temperature decreases monotonically leading
to a complete suppression of antiferromagnetic order at higher
doping accompanied by the disappearance of the pseudogap. The
absence of a pseudogap feature in PCO supports ideas, expressed in Refs.~%
\onlinecite{YKrock1, Sekitani, Naito}, about a strong suppression
or even absence of an antiferromagnetic insulating phase in
electron-doped cuprates, if the $T^\prime$ structure (i.e., no
apical oxygen) can be managed to survive down to very low or even
zero doping levels.

\section{Conclusions}

In our broadband investigation of the optical response of thin PCO
films, we unveiled a low-frequency (FIR) collective charge
excitation attributed to localization effects. As a function of
temperature, the optical spectral weight is redistributed between
this mode and the zero-frequency-centered Drude peak: the weight
of the FIR mode grows with temperature at the expense of the
Drude-peak weight. Such a behavior has been reported in underdoped
cuprates and is typical for bad metals.

We found that the optical spectral weight remains
temperature-dependent up to 1.9 eV, which indicates strong
electron correlations in PCO.

We calculated the electron-boson spectral density and found the
mass renormalization factor, $\lambda=4.16$ at $30\,$K.

In the millimeter-wave data, we directly observed the formation of
the superconducting condensate at $T < T_{c}$. We obtained that
the temperature dependence of the London penetration depth at low
temperatures follows a quadratic power law. This indicates
$d$-wave symmetry which is typical for the cuprates.

Neither the experimental optical data nor their analysis reveal
any indication of normal-state gap-like features which could be
attributed to the existence of a normal-state pseudogap. This
observation is in line with a breakdown of the
charge-transfer-insulator picture in PCO.

\section{Acknowledgements}

We are very grateful to Dr. Hideki Yamamoto for his work on sample
preparation and for useful discussions.

\appendix

\section{Thin Films Optical Conductivity}

\label{sec:ThinFlims}

In this work we measured a thin PCO film on top of a \DSco\
substrate. Harbecke\cite{Harbecke} studied the optical response of
multilayer systems in the most general case, including so-called
``coherent" and ``incoherent" light propagation.

Coherent propagation is expected when the optical thickness of a
medium is comparable or smaller than the wavelength of light. Our
measurements span wavelengths from the far-infrared to the
ultraviolet. The film thickness is of the order of a few hundred
angstroms. We can safely assume coherent propagation throughout
the whole measured range. Dealing with the substrate is trickier.
Our substrate is 0.5 mm thick. In the far-infrared, light
propagates coherently but coherence is gradually lost when moving
into the mid-infrared and shorter wavelengths.

To solve this problem, we worked with an unpolished back surface
of the substrate. The roughness of our back surface is of the
order of 10--50 $\mu$m, strongly diffusing light of shorter
wavelengths. For all that matters, light reflected from the back
surface will never reach the detector for this range. A problem
might still arise in the far-infrared. However, we have a highly
metallic, hence absorbing, film. The contribution from the back
surface of the substrate becomes negligible as it implies going
twice through the film and being partially diffused by the back
surface.

In practice, the effective geometry that best describes our system
is a thin film (thickness $d$ and complex refraction index $n_f$),
where light propagates coherently, sitting on top of a
half-infinite substrate having a complex refraction index $n_s$.
\cite{substrate} The near normal incidence reflectivity of such a
system is
\begin{equation}
R = \left|  r_0 + \frac{n_f t_0^2 r_f \phi_f^2}{1 + r_f r_0
\phi_f^2} \right|^2 \, , \label{ReflThinFilm}
\end{equation}
where $r_0=(1-n_f)/(a+n_f)$, $t_0 = 2 /(1+n_f)$, $r_f = (n_f -
n_s) / (n_f + n_s)$, and $\phi_f = \exp(2 \pi i \, n_f d /
\lambda)$, $\lambda$ being the wavelength of light.

In principle, one can perform Kramers-Kronig analysis on the
measured reflectivity and numerically invert
Eq.~\ref{ReflThinFilm} to obtain $n_f$, assuming that we measured
the substrate ($n_s$) independently. In practice this procedure is
strongly dependent on initial trial values and does not converge
with reasonable accuracy. An alternative approach is to model the
bulk dielectric function ($\varepsilon = n_f^2 =
4\pi\sigma/\omega$) of the film material. We enter this model into
Eq.~\ref{ReflThinFilm}, and utilize a least squares fitting to
refine its parameters.

The most straightforward modeling for the dielectric function is
the Drude-Lorentz formalism:
\begin{equation}
\varepsilon_{DL} = \varepsilon_\infty -
\frac{\omega_{pD}^2}{\omega^2 + i \omega \tau^{-1}} + \sum_k
\frac{\omega_{pk}^2}{\omega_{0k}^2 - \omega^2 - i \gamma_k
\omega}. \label{EpsDL}
\end{equation}
In Eq.~\ref{EpsDL}, $\varepsilon_\infty$ is the contribution from
electronic transitions in the deep-UV. The second term corresponds
to a free-carrier Drude response, characterized by a plasma
frequency $\omega_{pD}$, and a frequency-independent scattering
rate $\tau^{-1}$. The last term is a sum of Lorentz oscillators;
each of them is characterized by a resonance frequency
$\omega_{0k}$, a line width $\gamma_k$, and a plasma frequency
$\omega_{pk}$. All these parameters may freely vary with
temperature. This approach has been successfully utilized in
several cuprate thin films and a detailed analysis can be found in
Santander-Syro \textit{et al.}\cite{Santander}

However, the use of Eq.~\ref{EpsDL} is model dependent. We went a
step further in our analysis by adapting the procedure proposed to
extract optical functions from single crystal data by
Kuzmenko.\cite{Kuzmenko} In his approach the reflectivity is
roughly fitted by a Drude-Lorentz dielectric function. A second
step is then taken by choosing a variational dielectric function
$\varepsilon_V$. This function is added to $\varepsilon_{DL}$ and
adjusted in order to describe the total reflectivity within data
noise. In Kuzmenko's paper, $\varepsilon_V$ is obtained by setting
an arbitrary piece-wise imaginary part
($\varepsilon^{\prime\prime}$) of the dielectric function, and
calculating the corresponding real part from Kramers-Kronig. This
piecewise $\varepsilon^{\prime\prime}$ is obtained by simply
setting an arbitrary value at each measured frequency. This value
is modified in a least-squares fit so that the reflectivity is
properly adjusted. Kuzmenko showed that this method produces
optical functions with an accuracy equivalent to Kramers-Kronig.

In principle, Kuzmenko's method can be straightforwardly adapted
to the case of thin films if one replaces the equation for the
normal-incidence reflectivity in bulk materials by
Eq.~\ref{EpsDL}. However, because of the very large spectral range
of our data setting an arbitrary value for
$\varepsilon^{\prime\prime}_V$ at each point measured would be
impractical. What we did, instead, was to add a very large number
(of the order of 1000) of Lorentz oscillators distributed over the
regions where the first fit by the Drude-Lorentz model does not
describe properly the data. We fixed the frequency and width of
these oscillators and allowed its intensity to vary. In
particular, we modified the form shown in Eq.~\ref{EpsDL} in order
to allow for negative values of the numerator. These modified
Lorentz oscillators should not be regarded as physical excitations
of the systems. They represent a differential correction on the
rough Drude-Lorentz dielectric function.

At the end of the procedure, we have a dielectric function
\begin{equation}
\varepsilon = \varepsilon_{DL} + \varepsilon_V, \label{EpsEff}
\end{equation}
which is a model-independent description of the bulk optical
properties of the thin-film material. As a bonus, the rough
$\varepsilon_{DL}$ gives us an overall idea of the physical
excitations in the system. However, the obtained dielectric
function is model-independent and can be used even when the
Drude-Lorentz approach is not expected to work, such as in the
superconducting state. Furthermore, usage of the Drude term is not
a must even in a metallic state. One could use a large number of
Lorentz oscillators instead.

\section{Maximum entropy inversion}
\label{sec:MaxEnt}

The optical scattering rate $\tau^{-1}_{op}(\omega)$ is directly
related to the electron-exchange boson interaction spectral
density (Eliashberg function) $I^2\chi(\omega)$
via\cite{allen,ESchach1}
\begin{equation}
 \tau^{-1}_{op}(\omega,T)-\tau^{-1}_{imp} =
 \int\limits_0^\infty\!d\Omega\,K(\omega,\Omega;T)I^2\chi(\omega).
 \label{eq:InvTau}
\end{equation}
The kernel $K(\omega,\Omega;T)$ is determined from theory,
$\tau^{-1}_{imp}$ is an impurity scattering rate and
$I^2\chi(\omega)$ can be calculated by deconvoluting (inverting)
Eq.~\eqref{eq:InvTau}. Shulga {\it et al.}\cite{shulga} derived
for the normal state the kernel
\begin{eqnarray}
  K(\omega,\Omega;T) &=& \frac{\pi}{\omega}\left [
   2\omega\text{coth}\left(\frac{\Omega}{2\pi}\right)-
   (\omega+\Omega)\text{coth}\left(\frac{\omega+\Omega}{2T}\right)
   \right.\notag\\
  &&\left. +(\omega-\Omega)\text{coth}\left(\frac{\omega-\Omega}
   {2T}\right)\right].
  \label{eq:Shulga}
\end{eqnarray}
This kernel is a very good approximation to the exact result of
Eliashberg theory. It is valid for any temperature $T$ and reduces
to the kernel derived by Allen\cite{allen} for $T=0$.

The maximum-entropy method to invert Eq.~\eqref{eq:InvTau} to gain
information on $I^2\chi(\omega)$ from experimental
$\tau^{-1}_{op}(\omega,T)$ data was described in detail by
Schachinger {\it et al.}\cite{ESchach2} This process is, of
course, an ill-posed problem which does not necessarily allow for
an unique solution.

Let us define
\begin{equation}
  \label{eq:chisq}
  \chi^2 = \sum\limits_{i=1}^N\frac{[D_i-\tau^{-1}_{op}(\omega_i)]^2}
   {\varepsilon^2_i},
\end{equation}
where $D_i$ are the experimental $\tau^{-1}_{op}$ data points at
discrete energies $\omega_i$ and $\tau^{-1}_{op}(\omega_i)$ is
calculated from Eq.~\eqref{eq:InvTau} and kernel
Eq.~\eqref{eq:Shulga} and is to be regarded as a functional of
$I^2\chi(\omega)$. Finally, $\varepsilon_i$ denotes the error bar
on the data $D_i$ and $N$ is the number of data points.
Furthermore, physics requires that the spectral density
$I^2\chi(\omega)$ is positive definite. To achieve this the
maximum-entropy method minimizes the functional
\begin{equation}
  \label{eq:func}
  L = \frac{\chi^2}{2}-aS,
\end{equation}
with $S$ the generalized Shannon-Jones entropy,\cite{sivia}
\begin{equation}
  \label{eq:shannon}
  S = \int\limits_0^\infty\!d\omega\,\left\{I^2\chi(\omega)-m(\omega)-
I^2\chi(\omega)\ln\left[\frac{I^2\chi(\omega)}{m(\omega)}\right]
\right\},
\end{equation}
which gets maximized in the process. $m(\omega)$ is the constraint
function (default model) which reflects {\it a priori} knowledge
of $I^2\chi(\omega)$. In Eq.~\eqref{eq:func} $a$ is a
determinative parameter that controls how close the fitting should
follow the data while not violating the physical constraints. In
our inversion we set $m(\omega) = m_0$ for $\omega_1\le\omega\le
\omega_N$ with $m_0$ some small constant indicating that we have
no knowledge whatsoever about $I^2\chi(\omega)$, thus establishing
an {\it unbiased} inversion of Eq.~\eqref{eq:InvTau}.  Finally we
will make use of the historical maximum-entropy method which
iterates $a$ until the average $\langle\chi^2\rangle = N$ is
achieved with acceptable accuracy.


\begin{thebibliography}{99}

\bibitem{Imada} M. Imada, A. Fujimori, and Y. Tokura, Rev. Mod. Phys. \textbf{70}, 1039 (1998).

\bibitem{ARM} N. P. Armitage, P. Fournier, and R. L. Greene, Rev. Mod. Phys.
\textbf{82}, 2421 (2010).

\bibitem{Brinkmann} M. Brinkmann, T. Rex, H. Bach, and K. Westerholt,
Phys. Rev. Lett. \textbf{74}, 4927 (1995).

\bibitem{Matsumoto1} O. Matsumoto, A. Utsuki, A. Tsukada, H. Yamamoto,
T. Manabe, and M. Naito, Phys. Rev. B \textbf{79},  100508(R) (2009).

\bibitem{Matsumoto2} O. Matsumoto, A. Utsuki, A. Tsukada, H. Yamamoto,
T. Manabe, and M. Naito, Physica C \textbf{468}, 1148 (2008).

\bibitem{Matsumoto3} O. Matsumoto, A. Tsukada, H. Yamamoto, T. Manabe,
and M. Naito, Physica C \textbf{470}, 1029 (2010).

\bibitem{Matsumoto4} O. Matsumoto, A. Utsuki, A. Tsukada, H. Yamamoto, T. Manabe, and M.
Naito, Physica C \textbf{469}, 924 (2009).

\bibitem{Yamamoto} H. Yamamoto, O. Matsumoto, Y. Krockenberger, K. Yamagami,
and M. Naito, Solid State Commun. \textbf{151}, 771 (2011).

\bibitem{Ikeda} A. Ikeda, O. Matsumoto, H. Yamamoto, T. Manabe, and M. Naito, Physica C \textbf{471}, 686 (2011).

\bibitem{YKrock1} Y. Krockenberger, H. Irie, O. Matsumoto, K. Yamagami,
M. Mitsuhashi, A. Tsukada, M. Naito, and H. Yamamoto, Sci. Rep. \textbf{3}, 2235 (2013).

\bibitem{Sekitani} T. Sekitani, M. Naito, and N. Miura, Phys. Rev. B \textbf{67}, 174503 (2003).

\bibitem{Naito} M. Naito, O. Matsumoto, A. Utsuki, A. Tsukada, H. Yamamoto, and T.
Manabe, J. Phys.: Conf. Ser. \textbf{108}, 012037, (2008).

\bibitem{Das} H. Das and T. Saha-Dasgupta, Phys. Rev. B \textbf{79}, 134522 (2009).

\bibitem{Weber1} C. Weber, K. Haule, and G. Kotliar, Nat. Phys. \textbf{6}, 574 (2010).

\bibitem{Weber3} C. Weber, K. Haule, and G. Kotliar, Phys. Rev. B \textbf{82}, 125107 (2010).

\bibitem{Calder} Only recently, the first experimental evidence for realization of a
Slater transition has been reported: S. Calder, V. O. Garlea, D.
F. McMorrow, M. D. Lumsden, M. B. Stone, J. C. Lang, J.-W. Kim, J.
A. Schlueter, Y. G. Shi, K. Yamaura, Y. S. Sun, Y. Tsujimoto, and
A. D. Christianson, Phys. Rev. Lett. \textbf{108}, 257209 (2012).

\bibitem{Schultz} A. J. Schultz, J. D. Jorgensen, J. L. Peng,
and R. L. Greene, Phys. Rev. B \textbf{53}, 5157 (1996).

\bibitem{Radaelli} P. G. Radaelli, J. D. Jorgensen, A. J. Schultz,
J. L. Peng, and R. L. Greene, Phys. Rev. B \textbf{49}, 15322 (1994).

\bibitem{Ueda} S. Ueda, S. Asai, and M. Naito, Physica C \textbf{470}, 1173 (2010).

\bibitem{Asai} S. Asai, S. Ueda, M. Naito, Physica C \textbf{471}, 682 (2011).

\bibitem{TAdachi} T. Adachi, Y. Mori, A. Takahashi, M. Kato, T. Nishizaki,
T. Sasaki, N. Kobayashi, and Y. Koike, J. Phys. Soc. Jpn.
\textbf{82}, 063713 (2013).

\bibitem{gold} C. C. Homes, M. Reedyk, D. A. Crandles, and T. Timusk, Appl. Opt.
\textbf{32}, 2976 (1993).

\bibitem{Kozlov}  G. V. Kozlov and A. A. Volkov, in {\it Millimeter and
Submillimeter Wave Spectroscopy of Solids}, edited by G. Gr\"{u}ner
(Springer, Berlin, 1998), p. 51.

\bibitem{Dressel} M. Dressel, N. Drichko, B. P. Gorshunov, and A. Pimenov,
IEEE J. Sel. Top. Quantum Electron. \textbf{14}, 399 (2008) and
references therein.

\bibitem{Fischer} T. Fischer, A. V. Pronin, R. Skrotzki, T. Herrmannsd\"orfer, J. Wosnitza,
J. Fiedler, V. Heera, M. Helm, and E. Schachinger, Phys. Rev. B
\textbf{87}, 014502 (2013).

\bibitem{Tsuei} C. C. Tsuei, A. Gupta, and G. Koren, Physica C \textbf{161}, 415 (1989).

\bibitem{Onose} Y. Onose, Y. Taguchi, K. Ishizaka, and Y. Tokura,
Phys. Rev. B \textbf{69}, 024504 (2004).

\bibitem{Abrikosov} A. A. Abrikosov, L. P. Gorkov, and I. E. Dzyaloshinski,
\textit{Methods of Quantum Field Theory in Statistical Physics}
(Prentice-Hall, Englewood Cliffs, N.J., 1963).

\bibitem{Pines} D. Pines and P. Nozi\`{e}res, \textit{The Theory of Quantum Liquids} (Addison-Wesley, Reading, 1966), Vol. 1.

\bibitem{Tr_phase} Details of how the superconducting transition and the changes in
the temperature-dependent phase-sensitive optical measurements are
related, can be found in Ref. \onlinecite{Fischer}.

\bibitem{Homes} C. C. Homes, Q. Li, P. Fournier, and R. L. Greene,
Phys. Rev. B \textbf{66}, 144511 (2002).

\bibitem{Basov2} D. N. Basov, R. D. Averitt, D. van der Marel, and
M. Dressel, Rev. Mod. Phys. \textbf{83}, 471 (2011).

\bibitem{Emery} V. J. Emery and S. A. Kivelson, Phys. Rev. Lett. \textbf{74}, 3253 (1995).

\bibitem{Mutou} T. Mutou and H. Kontani, Phys. Rev. B \textbf{74}, 115107 (2006).

\bibitem{Takenaka} K. Takenaka, R. Shiozaki, S. Okuyama, J. Nohara,
A. Osuka, Y. Takayanagi, and S. Sugai, Phys. Rev. B \textbf{65},
092405 (2002).

\bibitem{Basov3} D. N. Basov, B. Dabrowski, and T. Timusk,
Phys. Rev. Lett. \textbf{81}, 2132 (1998).

\bibitem{Arima} T. Arima, Y. Tokura, and S. Uchida, Phys. Rev. B
\textbf{48}, 6597 (1993).

\bibitem{Lobo} A similar effect, although not as a function of temperature, but
rather of doping, was reported in: R. P. S. M. Lobo, E. Y.
Sherman, D. Racah, Y. Dagan, and N. Bontemps, Phys. Rev. B
\textbf{65}, 104509 (2002).

\bibitem{Qazilbash} M. M. Qazilbash, K. S. Burch, D. Whisler, D. Shrekenhamer,
B. G. Chae, H. T. Kim, and D. N. Basov, Phys. Rev. B \textbf{74},
205118 (2006).

\bibitem{Uchida} S. Uchida, T. Ido, H. Takagi, T. Arima, Y. Tokura, and
S. Tajima, Phys. Rev. B \textbf{43}, 7942 (1991).

\bibitem{Lee} Y. S. Lee, K. Segawa, Z. Q. Li, W. J. Padilla, M. Dumm,
S. V. Dordevic, C. C. Homes, Y. Ando, and D. N. Basov, Phys. Rev. B \textbf{72}, 054529 (2005).

\bibitem{JWAllen} J. W. Allen and J. C. Mikkelsen, Phys. Rev. B \textbf{15}, 2952 (1977).

\bibitem{Puchkov} A. V. Puchkov, D. N. Basov and T. Timusk, J. Phys.: Condens. Matter \textbf{8}
10049 (1996).

\bibitem{ESchach3} E. Schachinger and J. P. Carbotte, in: {\it
Models and Methods of High-$T_{c}$ Superconductivity: Some Frontal
Aspects}, edited by J. K. Srivastava and S. M. Rao, (Nova Science,
Hauppauge, NY, 2003), Vol. II, p. 73.

\bibitem{ESchach1} E. Schachinger, J. J. Tu, and J. P. Carbotte,
Phys. Rev. B \textbf{67}, 214508 (2003).

\bibitem{ESchach} E. Schachinger, C. C. Homes, R. P. S. M. Lobo, and J. P. Carbotte,
Phys. Rev. B \textbf{78}, 134522 (2008).

\bibitem{lee} W. Lee, D. Rainer, and W. Zimmermann, Physica C
\textbf{159}, 535 (1989).

\bibitem{Hwang}J. Hwang, E. Schachinger, J. P. Carbotte, F. Gao,
D. B. Tanner, and T. Timusk, \prl \textbf{100}, 137005 (2008).

\bibitem{Basov1} D. N. Basov and T. Timusk, Rev. Mod. Phys. \textbf{77}, 721 (2005).

\bibitem{Zimmers1} A. Zimmers, R. P. S. M. Lobo, N. Bontemps, C. C. Homes, M. C.
Barr, Y. Dagan, and R. L. Greene, Phys. Rev. B \textbf{70}, 132502
(2004).

\bibitem{Pronin} A. V. Pronin, T. Fischer, J. Wosnitza, A. Ikeda,
and M. Naito, Physica C \textbf{473}, 11 (2012).

\bibitem{Homes1} C. C. Homes, S. V. Dordevic, M. Strongin, D. A. Bonn,
R. Liang, W. N. Hardy, S. Komiya, Y. Ando, G. Yu, N. Kaneko, X.
Zhao, M. Greven, D. N. Basov, and T. Timusk, Nature \textbf{430}
539, (2004).

\bibitem{Zaanen} J. Zaanen, Nature \textbf{430}, 512 (2004).

\bibitem{Zaanen1} The empirical relation of Ref.~\onlinecite{Homes1}
also works for conventional superconductors in dirty limit, but
because of different reasons, as discussed in
Ref.~\onlinecite{Zaanen}.

\bibitem{Timusk} T. Timusk and B. Statt, Rep. Prog. Phys. \textbf{62}, 61 (1999).

\bibitem{Zimmers} A. Zimmers, J. M. Tomczak, R. P. S. M. Lobo,
N. Bontemps, C. P. Hill, M. C. Barr, Y. Dagan, R. L. Greene, A. J.
Millis, and C. C. Homes, Europhys. Lett. \textbf{70}, 225 (2005).

\bibitem{YKrock} Y. Krockenberger, H. Yamamoto, A. Tsukada, M. Mitsuhashi,
and M. Naito, Phys. Rev. B \textbf{85}, 184502 (2012).

\bibitem{Harbecke} B. Harbecke, Appl. Phys. B \textbf{39}, 165 (1986).

\bibitem{substrate} The validity of considering the substrate as a half-infinite
medium was also justified by our measurements of the bare
substrate. Multiple reflections within the substrate would reveal
themselves either by interference fringes or by a characteristic
upturn at frequencies above some 1000 -- 2000 cm$^{-1}$. None of
these effects were detected (see the inset of
Fig.~\ref{reflectivity}). Instead, the bare-substrate spectra were
accurately described by Eq.~\eqref{ReflThinFilm} and a number of
Lorentzians.

\bibitem{Santander} A. F. Santander-Syro, R. P. S. M. Lobo, N. Bontemps,
W. Lopera, D. Girat\'{a}, Z. Konstantinovic, Z. Z. Li, and H.
Raffy, Phys. Rev. B \textbf{70}, 134504 (2004).

\bibitem{Kuzmenko} A. B. Kuzmenko, Rev. Sci. Instrum. \textbf{76}, 083108 (2005).

\bibitem{allen} P. B. Allen, Phys. Rev. B \textbf{3}, 305 (1971).

\bibitem{shulga} S. V. Shulga, O. V. Dolgov, and E. G. Maksimov,
Physica C \textbf{178}, 266 (1991).

\bibitem{ESchach2} E. Schachinger, D. Neuber, and J. P. Carbotte,
Phys. Rev. B \textbf{73}, 184507 (2006).

\bibitem{sivia} See, for example, D. S. Sivia, {\it Data Analysis}
(Clarendon Press, Oxford, 1996).

\end{thebibliography}
\end{document}